\documentclass[conference]{IEEEtran}
\IEEEoverridecommandlockouts
\usepackage{cite}
\usepackage{amsmath,amssymb,amsfonts}
\usepackage{algorithmic}
\usepackage{graphicx}
\usepackage{textcomp}
\usepackage{xcolor}
\def\BibTeX{{\rm B\kern-.05em{\sc i\kern-.025em b}\kern-.08em
    T\kern-.1667em\lower.7ex\hbox{E}\kern-.125emX}}

\begin{document}

\title{Hierarchical Tree-structured Knowledge Graph For Academic Insight Survey \\
\thanks{*This work was supported by JSPS KAKENHI Grant
Number JP20H04295.}
}

\author{Jinghong Li, Phan Huy, Wen Gu, Koichi Ota, Shinobu Hasegawa\\
Japan Advanced Institute of Science and Technology, Japan, FPT University, HCMC, Vietnam\\
lijinghong-n@jaist.ac.jp, huypt24@fe.edu.vn, wgu@jaist.ac.jp, ota@jaist.ac.jp, hasegawa@jaist.ac.jp\\
}

\maketitle

\begin{abstract}
Research surveys have always posed a challenge for novice researchers who lack research training. These researchers struggle to understand the directions within their research topic and the discovery of new research findings within a short time. One way to provide intuitive assistance to novice researchers is by offering relevant knowledge graphs (\textbf{\textit{KG}}) and recommending related academic papers. However, existing navigation knowledge graphs mainly rely on keywords or meta information in the research field to guide researchers, which makes it difficult to clearly present the hierarchical relationships, such as inheritance and relevance between multiple related papers. Moreover, most recommendation systems for academic papers simply rely on high text similarity, confusing researchers as to why a particular article is recommended. They may lack the grasp of important information about the insight connection between ‘Issue resolved' and ‘Issue finding' that they hope to obtain. This study aims to support research insight surveys for novice researchers by establishing a hierarchical tree-structured knowledge graph that reflects the inheritance insight and the relevance insight among multiple academic papers on specific research topics to address these issues.
\end{abstract}


\section{Introduction}
\label{main}

In recent years, there has been significant progress in the digital environment for academic papers. Databases of academic papers, such as \textit{S2orc}\cite{S2ORC} and \textit{Unarxiv}\cite{Unarxiv}, have been continuously improving. The advent of large language models (\textbf{\textit{LLM}}) like \textit{Chatgpt} have opened up new possibilities for utilizing these extensive data sources in NLP tasks, including question-answering systems and automatic summarization for academic papers. This advancement enables researchers to conduct research surveys more efficiently using research tools and resources, such as \textit{ChatPDF} and \textit{SciSpace}, which have been developed based on \textit{ChatGPT} and specifically tailored for academic papers understanding\cite{CTBR}. Additionally, knowledge graphs (\textbf{\textit{KG}}) built on academic papers, such as \textit{Connected Papers}\cite{Connected-paper}, make it more convenient for researchers to locate desired papers through keyword search. However, for researchers who are new to an academic field or who are exploring a specific research topic, it is difficult to quickly observe, understand, and explore research directions based on large repositories of academic articles solely by simply reading academic papers without a clear target\cite{VPRAS}. This inefficient survey approach hampers the discovery of research direction and innovation. Therefore, relying solely on paper reading comprehension support software and existing paper recommendation systems may make it difficult to achieve the goal of gaining research insights. Providing an overview of possible research directions and branches within multiple papers on a specific topic may greatly facilitate a more efficient exploration of research topics. This process involves the challenging task of conducting an in-depth exploration of keywords, inheritance, relevance, and their applicability within specific research directions. To address these issues, the objective of this study is to create a knowledge graph for insights survey assistants from multiple academic papers on a specific research topic. It presents a tree structure that shows:

\textbf{(1)} From the origin of the research task, expand the citation inheritance and relevance associations . Here, ‘Inheritance' refers to the route of research evolution from the origin (history) of a research direction to the current stage, which consists of at least two or more academic papers. ‘Relevance' refers to the path composed of multiple academic papers that aim to solve similar academic issues, which is formed by the relevance chain. Unlike ‘Inheritance' depended on citation connection, relevance tends to link the similarity of existing academic issues to discover the research direction.

\textbf{(2)} Explore the relevance chain within similar research tasks to demonstrate relevant research points. It provides potential valuable assistance to researchers in understanding, exploring, and gaining insights into research topics. Additionally, it guides researchers in quickly identifying topic clues and directions for further research expansion. This study contributes to an in-depth exploration and visualization of research topics in the following ways:

\textbf{1. }Based on the existing academic paper repository, we performed secondary development to fit the requirements of the research insight survey. We integrate a dataset that includes higher-dimensional features, which reflect the citation structure among papers and enable insights content segmentation.

\textbf{2. } To explore the connections of academic issue relevance between pairs of papers, we utilize machine learning technology to build a training model for classifying the insight content into ‘Issue finding' and ‘Issue resolved'.

\textbf{3. } Based on the \textbf{(1)} and \textbf{2. }, we establish a \textbf{\textit{KG}} in a hierarchical tree structure. This \textbf{\textit{KG}} serves as a guide for novice researchers exploring research topics to achieve efficient insight surveys.

\section{Related work}

Research survey is a comprehensive investigation conducted to gain an overview of a research field or topic. Novice researchers need to comprehensively understand the knowledge structure in their field through research surveys. The development of \textbf{\textit{KG}} can benefit for novice researchers in efficiently comprehending and exploring research topics. As previous studies on the \textbf{\textit{KG}} generated from academic papers, Xu et al. constructed a PubMed knowledge graph that includes meta information such as bio-entities and authors\cite{Pubmed} and Martha et al. proposed the Tree of Science (ToS), which recommends articles based on their position in the graph of citation\cite{ToS}. However, for novice researchers, relying solely on meta or citation information is insufficient to provide enough information about the research content in that field to gain the overview concept map of the research branch. Deeper data mining from the academic paper is still required to refine the academic knowledge graph, which may provide high-quality guidance for the research survey. The knowledge graphs that involve internal information of academic papers include Chan et al. proposed representation ontology for a four-space integrated \textbf{\textit{KG}} (background, objectives, solutions, and findings) using NLP technology, as well as Tu et al. proposed Semantic Knowledge Graph \textbf{\textit{(SKG})} that integrates semantic concepts to represent the corpus\cite{literature-KG}\cite{SKG}. They both classify the content of papers into entity concepts and associated research key-points in their knowledge graphs. However, the association paths of complex concept maps can be cumbersome, making it difficult for novice researchers to accurately locate academic papers in a specific research direction, which may result in limited insight into the research branch. Thus, extracting insightful content from the paper is crucial to guide researchers in conducting effective surveys.
Hayashi et al. introduce disentangled paper summarization, where models generate both contribution and context summaries simultaneously to hint at the contribution overview of a specific research direction\cite{contribution-summarizing}. This type of summary may also link the contributions of multiple papers to perform a research survey of novelty in that research topic. However, the absence of summaries related to ‘issues that could be improved' makes it challenging for researchers to trace the evolution history in a particular research branch, such as the relevance of ‘existing issues' and ‘issues need to be resolved.' The above studies have yet to involve the intuitive and concise expression of the evolution trajectory and the implicit connection between multiple papers. Therefore, We are committed to mining the inheritance relationships among papers and exploring the relevant connections between their mentioned issues. We focus on expanding knowledge mining in the field of insight survey support, which aims to enable researchers to gain insights into research directions quickly. To achieve this, we propose a hierarchical tree-structured \textbf{\textit{KG}} based on the insight content of academic papers that reflects the inheritance relationships and relevance chain among academic papers.

\section{Methodology}

\begin{figure*}[htbp]
\centering
\fbox{
\includegraphics[width=14.6cm,height=5.6cm]{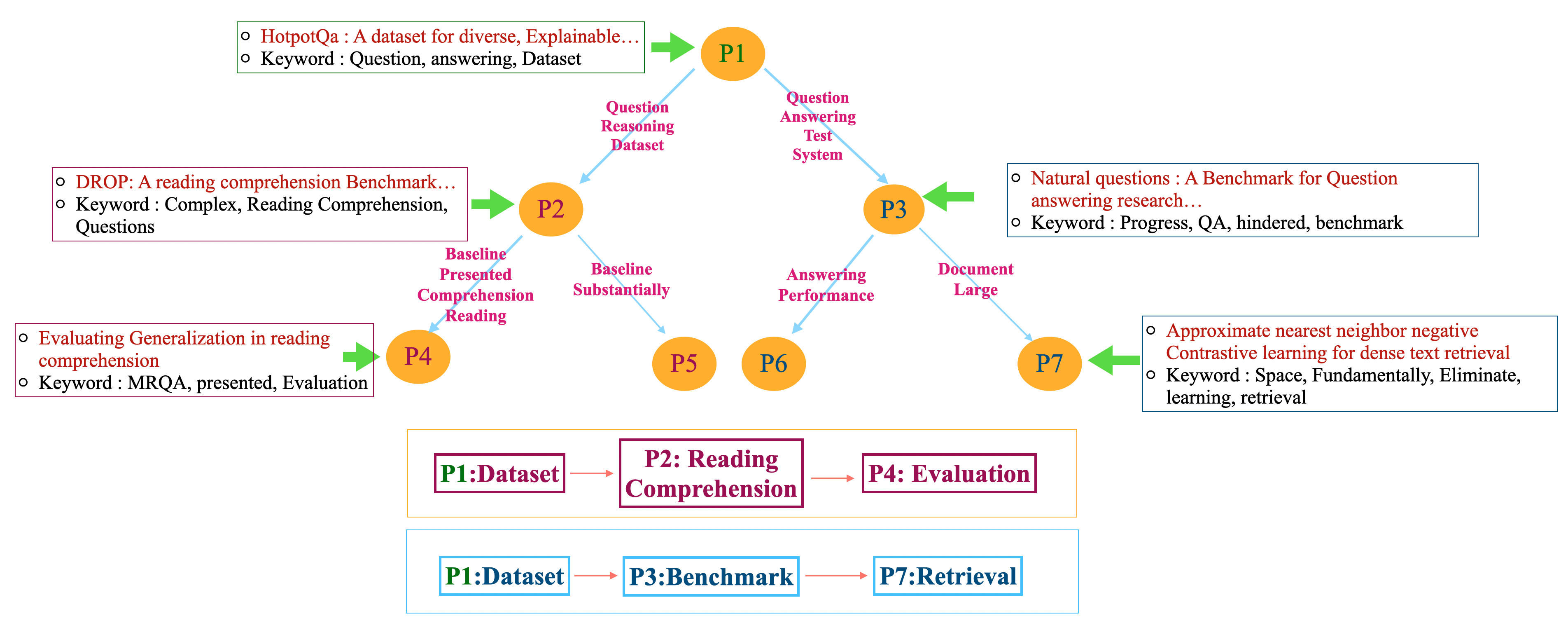}
}
\caption{Inheritance tree plan(‘HotpotQA topic')}
\label{figure1-1}
\end{figure*}

\begin{figure*}[htbp]
\centering
\fbox{
\includegraphics[width=14.6cm,height=5.6cm]{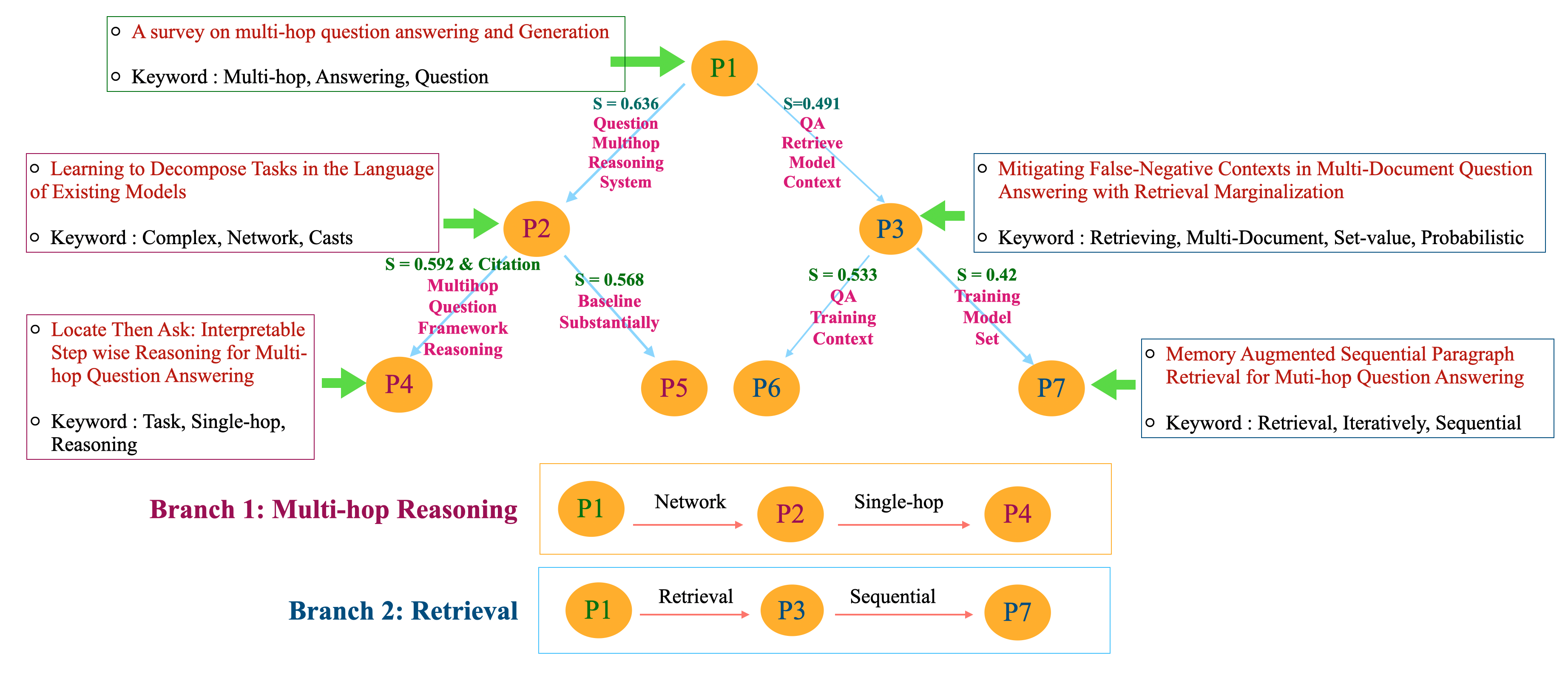}
}
\caption{Relevance tree plan(‘HotpotQA topic')}
\label{figure1-2}
\end{figure*}

In this work, we will use ‘HotpotQA' as an example, which is a well-known dataset in the field of NLP\cite{HotpotQA}. The Ideal \textbf{\textit{KG}} reflecting Inheritance insight and relevance insight survey of ‘HotpotQA' topic are shown in Fig.\ref{figure1-1}-\ref{figure1-2}. The configuration of the \textbf{\textit{KG}} is shown in Table \ref{table5}. In Fig.\ref{figure1-1}, the branches: \\
\verb|p1(Dataset)-> p2(Reading Comprehension)|\\ \verb|-> p4(Evaluation) and | \\ \verb|p1(Dataset)-> p3(Benchmark)-> p7(Retrieval)|
represent two research directions extended by the core paper of the ‘HotpotQA dataset'. This graph can help researchers infer the elements of research inheritance in the ‘HotpotQA' topic. For instance, \verb|P1| introduced the ‘HotpotQA dataset', \verb|P2| utilized this dataset for reading comprehension, and \verb|P4| developed evaluation metrics based on reading comprehension. In Figure \ref{figure1-2}, the path of ‘Multi-hop' branch: \\ \verb|p1 -> Network -> p2 -> Single hop -> p4| \\ and the path of ‘Retrieval' branch: \\ \verb|p1 -> Retrieval -> p3 -> Iteratively -> p7| \\ are extensions of the ‘Multi-hop' and ‘Retrieval' research tasks in the ‘HotpotQA' topic. This tree structure included multiple paths can help researchers infer the correlation factors between subtasks in the ‘HotpotQA' topic. For example, in the ‘Multi-hop' subtask, survey paper \verb|P1| provides a list of specific tasks, \verb|P2| extends it to the network establishment level, and \verb|P3| incorporates single-hop methods.

\begin{table*}[]
\centering
\caption{Knowledge graph configuration}
\label{table5}
\small
\begin{tabular}{lll}
\hline
\textbf{Property}        & \textbf{Content}                                                                                                                                                               & \textbf{Function}                                                                                                                            \\ \hline
\textit{Node$\_$label}     & Paper title                                                                                                                                                                    & Visually Presenting the theme of a Research Paper                                                                                            \\ \hline
\textit{Node$\_$Title}     & \begin{tabular}[c]{@{}l@{}}1. Keywords of paper \\ 2. ‘Issue Resolved' in this paper \\ 3. ‘Issue Finding' in this paper\end{tabular}                                              & \begin{tabular}[c]{@{}l@{}}Visually display the key information of research paper, \\ as a summary.\end{tabular}                             \\ \hline
\textit{Edge$\_$direction} & \begin{tabular}[c]{@{}l@{}}1. Inheritance tree :\\ Cite Source --\textgreater Cite Destination\\ 2. Relevance tree : \\ Issue Finding --\textgreater Issue Resolved\end{tabular} & \begin{tabular}[c]{@{}l@{}}1. Inheritance tree : Reflects the citation direction\\ 2. Relevance tree : Direction of Relevance chain\end{tabular} \\ \hline
\textit{Edge$\_$label}     & Co-occurring vocabulary                                                                                                                                                        & \begin{tabular}[c]{@{}l@{}}Helps user understanding of the \\ relevance between the two papers.\end{tabular}                                 \\ \hline
\textit{Edge$\_$Title}              & The value of relevance chain between papers                                                                                                                                    & Make it convenient for users to explore potential direction                                                                                  \\ \hline
\end{tabular}
\end{table*}

\subsection{Implementation procedure}
\begin{figure*}[htbp]
\centering
\includegraphics[width=15.0cm,height=7.3cm]{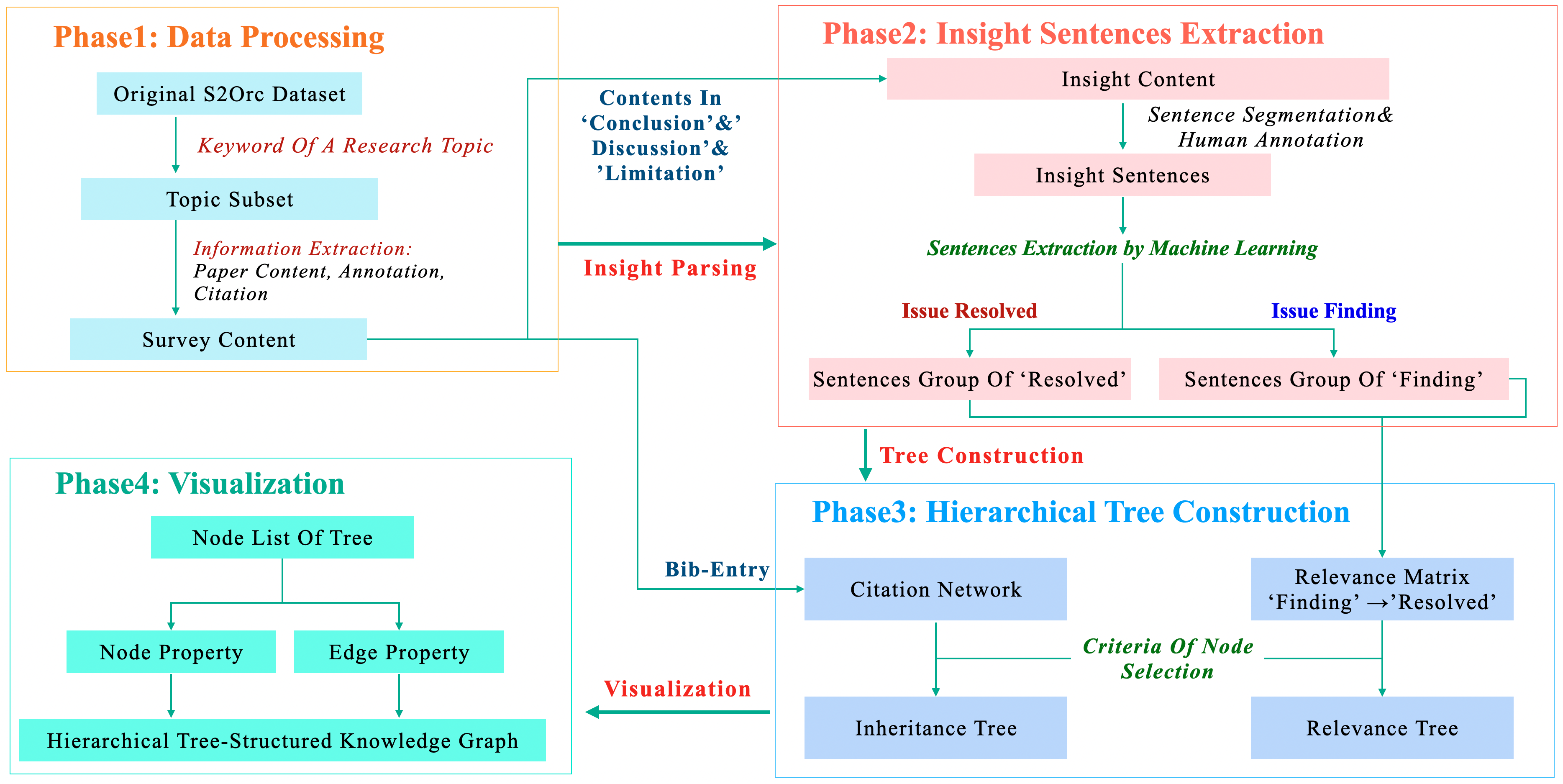}
\caption{Implementation procedure}
\label{figure_overview}
\end{figure*}

We utilize \textit{S2orc} as our infrastructure data of \textbf{\textit{KG}}. \textit{S2orc} is an academic database encompassing metadata, bibliographic references, and full text for 8.1 million open access papers\cite{S2ORC}. The overview of implementation is shown in Fig.\ref{figure_overview}. The development process is divided into 4 sub-stages:

\textbf{1.} In the first stage, we meticulously manipulate secondary data based on the \textit{S2orc} database to  construct our unique dataset tailored specifically for our insight survey. 

\textbf{2.} In the second stage, we use the \textit{Sentence Bert} model and manually set labels to perform three-class classification (Issue Resolved/Neutral/Issue Finding) on the insight content of each paper to extract the corresponding sentence.

\textbf{3.} In the third stage, we use the extracted sentences from phase 2 to analyze the inheritance and relevance chain in-depth. We select appropriate papers from the whole insight survey dataset based on certain criteria to generate tree-structured hierarchical trees.

\textbf{4.} In the final stage, we use the \textit{pyvis}\cite{pyvis} to visualize the tree-structured \textbf{\textit{KG}} and provide some case studies to demonstrate our findings. 

\subsection{Phase1 : Data processing}
To build the insight survey dataset, the first step is to select a topic and filter out the relevant sub-dataset from the \textit{S2orc} dataset. We iterate through each paper in the \textit{S2orc} dataset and extract those that contain the keyword ‘HotpotQA' into our sub-set. Our sub-set also includes data-citation information, full-text content, and meta-information of the papers associated with ‘HotpotQA.' Next, we extract the text from sections titled ‘conclusion',‘discussion' and ‘limitation' to identify insight content. We focus on paragraphs with section titles containing ‘conclusion', ‘discussion' and ‘limitation' as they address the problems discussed in the research paper and highlight any remaining challenges or limitations. Additionally, we extract relevant citation relationships from the \textit{S2orc} dataset to create a global citation network based on the annotations of ‘bib-entry' . The details of insight survey content as described in Table \ref{table1}.

\begin{table*}[]
\centering
\caption{Insight dataset processing (HotpotQA)}
\label{table1}
\begin{tabular}{lll}
\hline
\textbf{Annotation in S2orc}                                                   & \textbf{Description}                                           & \multicolumn{1}{c}{\textbf{Way of Processing}}                                                                                                                                                                                                                                                                  \\ \hline
\textit{title}                                                                 & Paper title                                                    & Use the offset annotation of the paper title to extract the title text.                                                                                                                                                                                                                                         \\ \hline
\textit{\begin{tabular}[c]{@{}l@{}}section-header \\ \&Paragraph\end{tabular}} & Section header                                                 & \begin{tabular}[c]{@{}l@{}}To extract paragraphs related to the viewpoints of ‘Finding' and \\ 'Resolved' issues, locate a section header that includes the words \\ ‘Conclusion', ‘Discuss', and ‘Limitation'. This ensure that the starting\\  point of the extracted paragraph is appropriate.\end{tabular} \\ \hline
\textit{bibentry}                                                              & \begin{tabular}[c]{@{}l@{}}Link cited \\ corpusid\end{tabular} & \begin{tabular}[c]{@{}l@{}}Find the corpusid of a paper in \textit{S2orc} that matches the cited paper's \\ corpusid in the ‘HotpotQA' subset.\end{tabular}                                                                                                                                                              \\ \hline
\end{tabular}
\end{table*}

\subsection{Phase2 : Insight Sentence Extraction}
In this section, we discuss the process of the insight survey dataset. First, we divide the text in the ‘insight-content' into sentences. We annotate each sentence with the label ‘Issue Resolved',‘Neutral', or ‘Issue Finding' corresponding to the viewpoints the sentence expresses. Next, we use the \textit{Sentence Bert}\cite{sentence-bert} to vectorize each sentence and adopt Support Vector Machines (\textbf{\textit{SVM}}) classifier to distinguish the corresponding label for each sentence. Finally, sentences with the same ‘Issue Resolved' and ‘Issue Finding' labels in each article form the ‘insight sentence' of that article.

\textbf{Sentence Segmentation:}
\textit{Spacy} is an open-source natural language processing library for Python that offers an API to access its machine learning trained methods and properties\cite{Spacy}. This work uses the pre-trained model in \textit{Spacy} to implement sentence segmentation. \textit{Spacy} provides a pre-trained English library called ‘\textit{en$\_$core$\_$sci$\_$lg}' which includes a default sentence segmenter\footnote{https://allenai.github.io/scispacy/}\footnote{https://www.tutorialspoint.com/perform-sentence-segmentation-using-python-spacy}. Any complex segmentation patterns that failed were manually fixed.

\textbf{Human Annotation:}
For the segmented sentences mentioned above, experts determine the viewpoint of each sentence based on its meaning. The viewpoints include 'Issue Resolved,' 'Neutral,' or 'Issue Finding.' Sentences that thoroughly analyze research methods without explicitly highlighting contributions will be considered ‘Neutral.'  Sentences discussing potential future works in a particular field without specificity will also fall into the `Neutral' category. 'Issue Resolved` sentences need at least a combination of contribution and experimental outcomes, regardless of their positive or negative outcomes. Sentences that ambiguously hint at trends and recommendations may be classified as ‘Issue Finding.' When categorizing these three viewpoints, the references, tables, and figures of the sentences are unchanged. Table 2 shows specific definitions, distinguishing criteria, and examples of these three labels. The dataset, which consists of insight sentences and labels, has been published on \footnote{https://www.kaggle.com/datasets/dannyleeakira/dataset-for-academic-novelty-insight-survey}

\begin{table*}[]
\centering
\caption{Human annotation strategy}
\label{tab:my-table}
\begin{tabular}{lll}
\hline
\textbf{Sentence}                                                                                                                                                                                              & \textbf{Description}                      & \textbf{Label(Issue)} \\ \hline
\begin{tabular}[c]{@{}l@{}}In this paper, we present textbrewer, a flexible pytorch-based\\ distillation toolkit for nlp research and applications.\end{tabular}                                               & The target of this paper               & Resolved           \\ \hline
\begin{tabular}[c]{@{}l@{}}Finally, we show that our model facilitates interpretability \\ by learning an explicit hierarchy of tasks based on the skills \\ they require.\end{tabular}                        & Things achieved in this paper        & Resolved         \\ \hline
\begin{tabular}[c]{@{}l@{}}For the fourth setting, we pre-train the model for 2 epochs \\ on the fever dataset, followed by 4 epochs on liar-plus, \\ the fine-tune on politihop for 4 epochs.\end{tabular}    & More detail of the method they did        & Neutral        \\ \hline
We have conducted a series of experiments.                                                                                                                                                                     & Not related to the insight viewpoints & Neutral        \\ \hline
\begin{tabular}[c]{@{}l@{}}Future directions of our work may include using git in \\ downstream nlp applications where the graph inductive \\ bias is necessary and dataset is scarce.\end{tabular}            & Future work of this paper                 & Finding       \\ \hline
\begin{tabular}[c]{@{}l@{}}More causal approaches such as amnesiac probing, which \\ directly intervene in the underlying model's representations,\\  may better distinguish between these cases.\end{tabular} & Limitation of this paper                  & Finding       \\ \hline
\end{tabular}
\end{table*}

\textbf{Training, Sentence Extraction and evaluation:}
We adopt \textit{‘scibert$\_$scivocab$\_$uncased'} pre-training model\footnote{https://huggingface.co/allenai/scibert$\_$scivocab$\_$uncased} \cite{scivocab}, which was trained comprising 1.14M full-papers and 3.1B tokens, was sourced from Semantic Scholar, for sentence vectorization. \textit{scibert$\_$scivocab$\_$cased} exhibits adaptability to both the corpus and domain, making it suitable for our training data. We selected \textbf{\textit{SVM}} for classifying the vectorized sentences due to their strong generalization performance. We also used 1500 labeled sentences for training and validation data. The training and test data details are shown in Table \ref{table3}. To obtain the optimal \textbf{\textit{SVM}} parameters, we use the grid search\cite{GS} method and find the best parameters to apply to the test. The classification accuracy evaluation is presented in Table \ref{table4}. 
The classification result shows that the ‘Issue Resolved' class has a higher F1 Score, as the larger amount of data might influence it. The 'Neutral' and 'Issue Finding' classes have lower F1 scores, indicating challenges in achieving both high precision and recall. 
Based on the results, we extracted insight sentences of the ‘Issue Resolved' and ‘Issue Finding' by combining the sentences with corresponding labels within each article.

\begin{table}[htbp]
\centering
\caption{Detail of issue-status dataset}
\label{table3}
\begin{tabular}{|l|c|c|}
\hline
                  & \textbf{Train} & \textbf{Test} \\ \hline
\textit{Issue Resolved}   & 532            & 165           \\ \hline
\textit{Neutral}  & 334            & 121            \\ \hline
\textit{Issue Finding} & 259            & 89            \\ \hline
\textit{Total}            & 1125            & 375           \\ \hline
\end{tabular}
\end{table}

\begin{table}[htbp]
\centering
\caption{Classification result of issue status}
\label{table4}
\begin{tabular}{|l|c|c|c|}
\hline
                  & \multicolumn{1}{l|}{\textbf{Precision}} & \multicolumn{1}{l|}{\textbf{Recall}} & \multicolumn{1}{l|}{\textbf{F1-score}} \\ \hline
\textit{Issue Resolved}   & 0.90                                   & 0.85                               & 0.88                                  \\ \hline
\textit{Neutral}  & 0.62                                    & 0.73                                 & 0.67                                   \\ \hline
\textit{Issue Finding} & 0.75                                    & 0.71                                 & 0.73                                   \\ \hline
\end{tabular}
\end{table}

\subsection{Phase3-4 : Hierarchical Tree Construction \&  Visualization}
In this section, we utilize the sentences extracted in phase 2 and the citation information obtained in phase 1 to comprehensively extract the insight characteristics of the papers. We then employ two strategies to select specific papers to construct a tree-structured network.

\textbf{Similarity calculation:}
To create the relevance tree, we use the classification results from Phase 2 to determine the elements of the relevance chain (‘Issue finding' → ‘Issue Resolved'). We then use embedded insight sentences in sentence transformers\cite{sentence-trans} with ‘\textit{scibert$\_$scivocab$\_$uncased}' model to calculate the cosine similarity between insight sentences labeled as ‘Issue Finding' and those labeled as ‘Issue Resolved.' Next, we iterate through all the insight sentences in the papers and calculate the relevance chain to generate the relevance matrix. Based on the values in the relevance matrix, we select papers to construct the relevance tree in the following section.

\textbf{Construction:}
We extract specific nodes from the insight survey dataset and establish a hierarchical tree-structured according to the following rules, Where \verb|N| represents the maximum number of root papers (The number of trees in a \textbf{\textit{KG}}), \verb|n| is the root sequence of the selected paper, \verb|M| represents the maximum number of leaves, \verb|m| is the leaf sequence of the extracted paper, \verb|T| represents the maximum depth of the tree, and \verb|t| is the current depth of the tree.

\textbf{(1) Inheritance tree:}

\textit{\textbf{Step 1 - Root node determine:}} Sort all the papers in descending order based on the number of other papers that have cited them. Select the top \verb|N| papers with the highest citation counts as the root. This operation ensures that the root node has a high level of inheritability throughout the paper library.

\textit{\textbf{Step 2 - Leaf selection:}} From the candidate group of papers that cite the root node paper, select the top \verb|m| papers with the highest citation counts as the leaf nodes of root \verb|n|.

\textit{\textbf{Step 3 - Parent node update:}} Set each leaf node as a new root node and repeat Step 2. If a root node is not cited (unable to generate a leaf), the branch is terminated.\\

\textbf{(2) Relevance tree:}

\textit{\textbf{Step 1 - Root node determine:}} Calculate the average similarity scores for each paper in the relevance matrix. Select the top \verb|N| papers with the highest similarity score as the root nodes. This operation ensures that the root node has a high correlation index throughout the paper library.

\textit{\textbf{Step 2 - Leaf selection:}} From the candidate group of papers that have relevance chain with the root node paper. Select the top \verb|m| papers with the highest similarity score as the leaf nodes of root \verb|n|.

\textit{\textbf{Step 3 - Parent node update:}} Set each leaf node as a new parent node and repeat Step 2. If a root node does not have a relevance chain (unable to generate a leaf), the branch is terminated.\\

\textbf{(3) Common Rules: }

\textit{\textbf{Rule 1:}} The selected nodes cannot be selected again.

\textit{\textbf{Rule 2:}} Repeat Steps 1-3 until reaching the upper limit of \verb|N| or the maximum depth limit value \verb|T|. The network is then generated.\\


\section{Conclusion}
This study developed two types of hierarchical tree-structured knowledge graphs called inheritance tree and relevance tree to support insight surveys for research beginners. The process consists of four stages: data processing, insight sentence extraction, hierarchical tree construction, and \textbf{\textit{KG}} visualization. 
First, we conducted high-dimensional secondary development of the \textit{S2orc} dataset to create an insight survey dataset that includes citation information and insight content. Then, we extracted sentences from the insight survey dataset that express insight viewpoints on Issue finding' and 'Issue resolved' using machine learning techniques. These sentences were parsed and used to construct a relevance matrix. Finally, based on the citation information and relevance matrix, we generated two types of hierarchical tree structures: Inheritance and Relevance tree. The generated \textbf{\textit{KG}} demonstrates their rationality, indicating that they can provide key-information to assist researchers in gaining insights into the directions of the research topic. The knowledge graph also exhibits interpretability and potential for further development. For the future expansion and improvement of this study, the following points are proposed:

\textbf{(1) }Incorporate Additional Text for Text Similarity and Relevance Chain Computation: Currently, this study only utilizes the content from the ‘conclusion', ‘discussion', and ‘limitation' sections. However, the sections ‘abstract' and ‘introduction,' specifically the part discussing previous issues, contain insight elements related to ‘issue need to be solved in previous research' and the ‘objective in this study'. Therefore, integrating these texts for high-dimensional training can optimize the content in the knowledge graph.

\textbf{(2) }We objectively evaluated the classification accuracy of the 'Issue Finding' and 'Issue Resolved' viewpoints. However, a future challenge is incorporating researchers' subjective evaluations into the generated knowledge graph.

\clearpage


\begin{thebibliography}{}

\bibitem{S2ORC}
Kinney, R., Anastasiades, C., Authur, R., Beltagy, I., Bragg, J., Buraczynski, A., ... \& Weld, D. S. (2023). The semantic scholar open data platform. arXiv preprint arXiv:2301.10140.


\bibitem{Unarxiv}
Saier, T., Krause, J., \& Färber, M. (2023). unarxive 2022: All arxiv publications pre-processed for nlp, including structured full-text and citation network. arXiv preprint arXiv:2303.14957.

\bibitem{Connected-paper}
Ammar, W., Groeneveld, D., Bhagavatula, C., Beltagy, I., Crawford, M., Downey, D., ... \& Etzioni, O. (2018). Construction of the Literature Graph in Semantic Scholar. In Proceedings of NAACL-HLT (pp. 84-91).

\bibitem{CTBR}
Li, J., Gu, W., Ota, K. et al. Object Recognition from Scientific Document Based on Compartment and Text Blocks Refinement Framework. SN COMPUT. SCI. 5, 816 (2024). https://doi.org/10.1007/s42979-024-03130-7

\bibitem{VPRAS}
Li, J., Tanabe , H., Ota, K., Gu , W., \& Hasegawa , S. (2023). Automatic Summarization for Academic Articles using Deep Learning and Reinforcement Learning with Viewpoints. The International FLAIRS Conference Proceedings, 36(1). https://doi.org/10.32473/flairs.36.133308.

\bibitem{Pubmed}
Xu, J., Kim, S., Song, M., Jeong, M., Kim, D., Kang, J., ... \& Ding, Y. (2020). Building a PubMed knowledge graph. Scientific data, 7(1), 205.




\bibitem{ToS}
Zuluaga, M., Robledo, S., Arbelaez-Echeverri, O., Osorio-Zuluaga, G. A., \& Duque-Méndez, N. (2022). Tree of Science - ToS: A Web-Based Tool for Scientific Literature Recommendation. Search Less, Research More!. Issues in Science and Technology Librarianship, (100). https://doi.org/10.29173/istl2696

\bibitem{literature-KG}
Chen, H., \& Luo, X. (2019). An automatic literature knowledge graph and reasoning network modeling framework based on ontology and natural language processing. Advanced Engineering Informatics, 42, 100959.

\bibitem{SKG}
Tu, Y., Qiu, R., \& Shen, H. W. (2023). SKG: A Versatile Information Retrieval and Analysis Framework for Academic Papers with Semantic Knowledge Graphs. arXiv preprint arXiv:2306.04758.






\bibitem{contribution-summarizing}
Hayashi, H., Kryściński, W., McCann, B., Rajani, N., \& Xiong, C. (2023, May). What’s New? Summarizing Contributions in Scientific Literature. In Proceedings of the 17th Conference of the European Chapter of the Association for Computational Linguistics (pp. 1019-1031).

\bibitem{HotpotQA}
Yang, Z., Qi, P., Zhang, S., Bengio, Y., Cohen, W., Salakhutdinov, R., \& Manning, C. D. (2018). HotpotQA: A Dataset for Diverse, Explainable Multi-hop Question Answering. In Proceedings of the 2018 Conference on Empirical Methods in Natural Language Processing (pp. 2369-2380).

\bibitem{sentence-bert}
Reimers, N., \& Gurevych, I. (2019, November). Sentence-BERT: Sentence Embeddings using Siamese BERT-Networks. In Proceedings of the 2019 Conference on Empirical Methods in Natural Language Processing and the 9th International Joint Conference on Natural Language Processing (EMNLP-IJCNLP) (pp. 3982-3992).

\bibitem{Spacy}
Ahmad, R., Shaikh, Y., \& Tanwani, S. (2023). Aspect Based Sentiment Analysis and Opinion Mining on Twitter Data Set Using Linguistic Rules. Indian Journal of Science and Technology, 16(10), 764-774.

\bibitem{scivocab}
Beltagy, I., Lo, K., \& Cohan, A. (2019, November). SciBERT: A Pretrained Language Model for Scientific Text. In Proceedings of the 2019 Conference on Empirical Methods in Natural Language Processing and the 9th International Joint Conference on Natural Language Processing (EMNLP-IJCNLP) (pp. 3615-3620).

\bibitem{GS}
Syarif, I., Prugel-Bennett, A., \& Wills, G. (2016). SVM parameter optimization using grid search and genetic algorithm to improve classification performance. TELKOMNIKA (Telecommunication Computing Electronics and Control), 14(4),1502-1509

\bibitem{tfidf-org}
Aizawa, A. (2003). An information-theoretic perspective of tf–idf measures. Information Processing \& Management, 39(1), 45-65

\bibitem{tfidf-keyword}
Ao, X., Yu, X., Liu, D., \& Tian, H. (2020, June). News keywords extraction algorithm based on TextRank and classified TF-IDF. In 2020 International Wireless Communications and Mobile Computing (IWCMC) (pp. 1364-1369). IEEE.

\bibitem{sentence-trans}
Reimers, N., \& Gurevych, I. (2019, November). Sentence-BERT: Sentence Embeddings using Siamese BERT-Networks. In Proceedings of the 2019 Conference on Empirical Methods in Natural Language Processing and the 9th International Joint Conference on Natural Language Processing (EMNLP-IJCNLP) (pp. 3982-3992).

\bibitem{pyvis}
Perrone, G., Unpingco, J., \& Lu, H. M. (2020). Network visualizations with Pyvis and VisJS.

 \end{thebibliography}
\end{document}